\documentclass[12pt]{article}
\usepackage[dvips]{graphicx}
\usepackage[american]{babel}
\usepackage{amsmath}

\begin{document}
\title{Statistical considerations on safety analysis}
\author{L. P\'al and M. Makai, \\\footnotesize{KFKI Atomic Energy Research Institute
1525 Budapest, P.O.B. 49. Hungary}}
\date{November 17, 2005}

\maketitle

\begin{abstract}
Alerting experience with a well-acknowledged safety analysis code
initiated the authors to pay attention to safety issues of complex
systems. Their first concern was the statistical characteristics
of such a code. We point out a remarkable weakness of the so
called 0.95/0.95 methodology: when repeating the search for the
tolerance limit, we get a higher value with non-negligible
probability. We propose the sign test as an alternative method. We
point out the correct form of Wilks' formula when the number of
parameters subjected to limitation is two or more.

\vspace{0.2cm}


\end{abstract}

\section{Introduction}

Alerting experience with a well-acknowledged safety analysis code
\cite{burwell89}, \cite{aust98} which is widely used in the
licensing process of nuclear power plants, initiated the authors
to pay attention to safety issues of nuclear reactors. Their first
concern was the statistical characteristics of such a code. In
order to judge if a given nuclear reactor was safe, one had to
demonstrate that safety criteria are met with a reasonable
probability. But to judge the output of the code, one needed to
know the probability distribution of the output.

In a former paper \cite{guba03}  we discussed the handling of
statistics of model calculations with several outputs. The present
work provides a correct statistical estimation of a quantile and
we point out the inadequacy of the traditional 95{\%} probability
limit approach, which seems to be the practice at US Nuclear
Regulatory Commission. We advocate the sign test instead.

Let us consider results of $N$ runs of a code modelling the single output variable,
which is subjected to  limitation. Let the output values be ordered:
\begin{equation} \label{1}
y(1) < y(2) < \cdots < y(N).
\end{equation}
We call the ensemble (\ref{1}) a sample. Let the acceptance range
be given as $(-\infty, U_{T}]$, where $U_{T}$ is the technological
limit for $y$. We assume that the distribution of $y$ is unknown,
and are looking for a quantile $Q_{\gamma}$ such that
\begin{equation} \label{2}
\int_{-\infty}^{Q_{\gamma}} dG(y) = \gamma,
\end{equation}
where $G(y)$ is the unknown cumulative distribution function of
output variable $y$. Quantile $Q_{\gamma}$ is to be derived from
measured value, thus, itself is a random variable.

In Section 2, we address the problem of estimating quantile $Q_\gamma$. Two solutions
are mentioned: the classical Baysian solution and a recent solution, which is applicable
to several variables. In Section 3, we present an example where the 0.95|0.95
methodology seems to fail and in Section 4, we suggest another methodology based on sign
test. Our concluding remarks are summarized in the last Section.

\section{Estimation for one-tailed quantile}

The random interval $(-\infty, y(s)]$ covers a proportion larger
than $\gamma$ of the unknown distribution function $G(y)$ with
probability $\beta$ when
\begin{equation} \label{3}
\beta = {\mathcal P}\{y(s) > Q_{\gamma}\},
\end{equation}
where ${\mathcal P}\{{\mathcal A}\}$ denotes the probability of
event $\mathcal A$. It can be shown [4] that
\begin{equation} \label{4}
\beta = \sum_{j=0}^{s-1} \binom{N}{j}\;
\gamma^{j}\;(1-\gamma)^{N-j}.
\end{equation}
When $s=N$, i.e. the largest element of the sample is chosen as
upper limit of the random interval, one obtains the well-known
formula:
\begin{equation} \label{5}
\beta  = 1 - \gamma^{N}.
\end{equation}
Since  one finds misinterpretations in the engineering practice it
is not superfluous to underline the proven notion of formula
(\ref{5}). $\beta $ is the probability that the largest value
$y(N)$ of a sample comprising $N$ observations is greater then the
$\gamma$ quantile of the unknown distribution of output variable
$y$. Another formulation asserts that $\beta $ is the probability
that the interval $(-\infty,y(N)]$ covers a larger than $\gamma$
portion of the unknown distribution $G(y)$ of the output variable
$y$.

\subsection{Old Bayesian method}

If we carry out $N$ runs, i.e., we determine the output variable from $N$ fluctuating
inputs, and define a fix acceptance region ${\mathcal H}_{a} = [L_{T}, U_{T}]$. The
probability
\[ {\mathcal P}\{y \in {\mathcal H}_{a}\} =
\int_{{\mathcal H}_{a}} g(u)\;du = w \]

\begin{table}[ht!]
\caption{\label{tab1} {\footnotesize Number of failures
observations $N-k$ at which $w \geq \omega$ holds with probability
at least $\alpha$}}
\begin{center}
\begin{tabular}{|c|c|c|c|c|} \hline
$\alpha$  & $\omega$  & $N-k=0$ & $N-k=1$ & $N-k=2$ \\ \hline
     & 0.90 &   21  &   31  &   51 \\
0.90 & 0.95 &   44  &   75  &  104 \\
     & 0.99 &  228  &  387  &  530 \\ \hline
     & 0.90 &   27  &   45  &   60 \\
0.95 & 0.95 &   57  &   92  &  123 \\
     & 0.99 &  297  &  472  &  626 \\ \hline
     & 0.90 &   43  &   63  &   80 \\
0.99 & 0.95 &   89  &  129  &  164 \\
     & 0.99 &  457  &  660  &  836 \\ \hline
\end{tabular}
\end{center}
\end{table}
\noindent of the output variable $y$ to lay in ${\mathcal H}_{a}$
is unknown. However, knowing that $k$ elements out of $N$ are in
the acceptance interval, we can estimate the probability that the
unknown acceptance probability $w$ is greater than a prescribed
$\omega$ without knowing the distribution function $g(u)$. The
claim is based on Bayes theorem on conditional probabilities and
asserts
\begin{equation} \label{6}
\beta(\omega\vert N,k) = \sum_{j=0}^{k}
\binom{N+1}{j}\;(1-\omega)^{j}\;\omega^{N + 1 - j}.
\end{equation}
\noindent The proof is available in textbooks. Using (\ref{6}), we
can easily determine the allowed number of rejections in a sample
of $N$ elements to make sure that $w \geq \omega$ is true with a
given $\beta \geq \alpha $ prescribed probability. In Tab. 1, we
have collected a few examples to give an impression how expression
(\ref{6}) works. It is noteworthy that even if $k=0$, i.e. when
all outputs are accepted, there is a non-zero probability that
outputs will appear which should have been rejected. As we see, no
failure out of 21 runs assures the same probability as one failure
out of 31 runs or two failures out of 51 runs (cf. the first row
of Tab. 1).

\subsection{Case of Several Variables}

The following statement generalizes the estimate of a quantile to several output
variables. In the case of $n \ge 2$ output variables with continuous joint distribution
function $G(y_{1}, \ldots, y_{n})$ it is possible to construct $n$-pairs of random
intervals $[L_{j},U_{j}],\; j=1,\ldots, n$ such that the probability of the inequality
\begin{equation} \label{7}
\int_{L_{1}}^{U_{1}} \cdots \int_{L_{n}}^{U_{n}} g(y_{1}, \ldots,
y_{n})\;dy_{1} \cdots dy_{n} > \gamma
\end{equation}
is free of $g(y_{1}, \ldots, y_{n})$ and is given by \[{\mathcal
P}\left\{\int_{L_{1}}^{U_{1}} \cdots \int_{L_{n}}^{U_{n}} g(y_{1},
\ldots, y_{n})\;dy_{1} \cdots dy_{n}
> \gamma \right\} =  \beta, \]
were $0 < \beta \le 1$ is a given number. Details and proof of the
statement can be found in [4].

\section{Challenge of the $\mathbf{0.95 \vert 0.95}$ methodology}

In the present section, we consider an example. We assume the
single output variable $y$ to have a lognormal distribution with
parameters $m$ and $d$. This will be our "unknown"  $G(y)$
distribution. The density function is
\begin{equation} \label{8}
g(y) = \frac{1}{yd \sqrt {2\pi}} \exp \left[- \frac{1}{2} \left(
\frac{\log y - m}{d}\right)^{2} \right],
\end{equation}
where $y \ge 0$.

We use Monte Carlo simulation to generate four samples of size
$N=100$, in the simulation we take $m=2.5$ and $d=0.5$. The goal
is to get point estimates of $0.95$-quantiles for each sample and
to determine the shortest two-tailed confidence intervals which
cover with $0.95$ probability the "unknown" quantile $Q_{0.95}$,
the reference value is $Q_{0.95} \approx 27.73$. The four samples
are labeled as A, B, C, and D, the results of the simulation are
summarized in Tab. 2.

\begin{table}[ht!]
\caption{ \label{tab2} {\footnotesize Confidence intervals $[y(r),
y(s)]$ covering the quantile $Q_{0.95}$.}}
\begin{center}
\begin{tabular}{|c|c|c|c|c|} \hline
\mbox{ }  & $A$  & $B$ & $C$ & $D$  \\ \hline $y(r)$ & 22.66 &
25.21 & 22.48 & 23.29 \\  \hline
 $Q_{0.95}$ & {\it 27.73} & {\it 27.73} & {\it  27.73} & {\it 27.73}  \\  \hline
 $y(s)$ & 33.25 & 38.28 & 35.88 & 53.05  \\  \hline
 $(r,s)$ & (91, 100) & (91, 100) & (91, 100) & (91, 100) \\ \hline
\end{tabular}
\end{center}
\end{table}
If the upper limit, determined by the technology would be
U$_{T}$=40, then, cases A,B, and C could be considered only as
safe.

Setting $\beta=0.95$ and $\gamma=0.95$, from Eq. (\ref{5}) we get the sample size
$N=58$, i.e. the largest element of a sample having $58$ elements~\footnote{In the
practice $N=59$ is used.} should be chosen as $Q_{0.95}$. We performed the following
numerical experiment: Generated a sample of $58$ elements, that sample is called
\textit{basic sample}, in notation: $y^{(b)}$. Then, we repeat the sample generation
$n=1000$ times, thus obtaining the samples
\begin{figure} [ht!]
\centering{
\includegraphics[height=5cm, width=7.5cm]{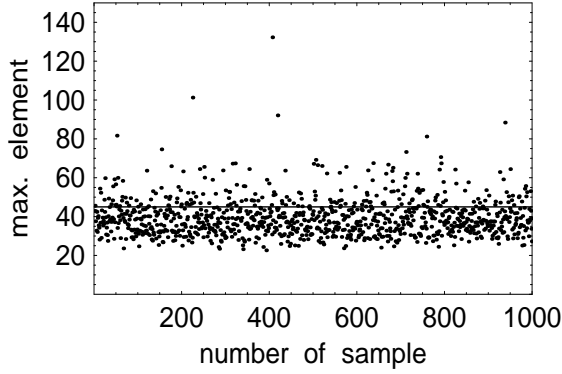}}
\caption{\label{fig1}{\footnotesize Results of of $1000$ samples
of size $N=58$. The largest element of the basic sample is
$y^{(b)}(58) \approx 45$.}}
\end{figure}
\noindent $y^{(1)}, y^{(2)},\ldots, y^{(1000)}$. The largest elements of those samples
can be seen in Fig. \ref{fig1}. The minimum of the values is $22.62$, the largest value
is $132.27$. One can observe that in $224$ samples (more than $22${\%} of the one
thousand samples) the maximum exceeds the maximum of the basic sample
($y^{(b)}(58)=45$). Let us check whether that number is reliable or not.

The probability that the largest element in a given sample is
greater than $Q_{\gamma}$ is $1-\gamma^{N}$. Let $\xi
_{n}(Q_{\gamma})$ stand for the random variable giving the number
of maximum elements exceeding $Q_{\gamma}$. The probability
distribution of the newly introduced random variable is
\begin{equation} \label{9}
{\mathcal P}\{\xi_{n}(Q_{\gamma}) = k\} = \binom{n}{k}
\;(1-\gamma^{N})^{k}\;\gamma^{N(n-k)}.
\end{equation}
From this expression we obtain the expectation value and the
variance as
\begin{equation} \label{10}
{\mathbf E}\{\xi_{n}(Q_{\gamma})\} = n(1-\gamma^{N}),
\end{equation}
\begin{equation} \label{11}
{\mathbf D}^{2}\{\xi_{n}(Q_{\gamma})\} =
n\;\gamma^{N}\;(1-\gamma^{N}).
\end{equation}
When $n$ and $k$ are sufficiently large, the distribution of the
random variable
\begin{equation} \label{12}
\chi_{n}(Q_{\gamma}) = \frac{\xi_{n}(Q_{\gamma}) - {\bf
E}\{\xi_{n}(Q_{\gamma})\}}{{\bf D}\{\xi_{n}(Q_{\gamma})\}}
\end{equation}
is approximately standard normal, hence,
\[ {\mathbf E}\{\xi_{n}(Q_{\gamma})\} - \lambda\;{\mathbf
D}\{\xi_{n}(Q_{\gamma})\} \leq \xi_{n}(Q_{\gamma}),\]
\begin{equation} \label{13}
\end{equation}
\[{\mathbf E}\{\xi_{n}(Q_{\gamma})\} + \lambda\;{\mathbf
D}\{\xi_{n}(Q_{\gamma})\} \geq \xi_{n}(Q_{\gamma}) \] is valid
with probability $w$ and $\lambda $ is the root of
\begin{equation} \label{14}
\frac{1}{\sqrt{2\pi}}\;\int_{-\infty}^{\lambda} e^{-u^{2}/2}\;du =
\frac{1+w}{2}.
\end{equation}
Substituting here $n=1000$, $N=58$ and $w=0.95$, we get
$\mathbf{E}\{\xi_{n}(Q_{\gamma})\}=950, \; \mathbf{D}\{\xi
_{n}(Q_{\gamma})\} \approx 6.96, \; \lambda \approx 1.96$, and the
following relationship is fulfilled with probability $95${\%}:
$936 < \xi_{1000}(Q_{0.95}) < 964$. We can not estimate the number
of samples, in each of which the maximum exceeds the maximum of
the basic sample but we can count the number of maximal values
exceeding the known quantile $Q_{0.95}$, that number is 949, a
number witnessing the correctness of the statistics.

In spite of the nice agreement we wish to underline that the
($0.95 \vert 0.95$) safety policy does not exclude rare events
such as limit violation when some of the calculated values are
over the limit $U_{T}$.

\textbf{Another conclusion is that the maximal element of a single sample of
$\mathbf{58}$ elements would be $\mathbf{y^{(b)}(58)}$ and if we repeat the sampling
several times, then in relatively large number of the samples we get a higher than
$\mathbf{y^{(b)}(58)}$ value for the maximal element. In the light of this experience
one asks: is this the intended outcome of the $\mathbf{0.95 \vert 0.95}$ methodology? It
is clear that a larger safety margin is needed to compensate for the weakness of the
$\mathbf{0.95 \vert 0.95}$ methodology.}

One must  mention here that the result found in the above presented example is not
exceptional but it is a direct consequence of a well-known theorem of mathematical
statistics. It is easy to show that if one repeats the sampling from any continuous
distribution $(n+1)$ times independently, then the probability that at least $k$ out of
$n$ maximal sample elements $y^{(1)}(N), \ldots, y^{(n)}(N)$ will exceed the initial
(basic) sample value $y^{(b)}(N)$, is equal to $1-k/(n+1)$. The proof of the theorem and
two important remarks are given in the Appendix.

\section{Method based on sign test}

The concluding remarks at the end of the previous section are not
optimistic. The question is whether one can find a method more
suitable for checking, from a computer model, the safety of a
large device? Below we propose such a method based on sign test.

Again, we assume the cumulative distribution function $G(y)$ of
the output variable to be continuous but unknown. Let
$S_{N}=\{y_{1},\ldots, y_{N}\}$ be a sample of $N$ observations
(runs of a computer model). Define the function
\begin{equation} \label{15}
\Delta(x) = \left\{ \begin{array}{ll}
1, & \mbox{if $x > 0$,} \\
\mbox{} & \mbox{} \\
0, & \mbox{if $x < 0$,} \end{array} \right.
\end{equation}
and introduce the statistical function
\begin{equation} \label{16}
z_{N} = \sum_{j=1}^{N} \Delta(U_{T} - y_{j})
\end{equation}
which gives the number of sample elements smaller than $U_{T}$.
Criteria based on statistical function (16) are called sign
criterion since $z_{N}$ counts the positive $U_{T} - y_{j}$
differences. When $G(y)$ is continuous, the probability of $U_{T}
- y = 0$ is zero.

Obviously, distribution of $z_{N}$ is binomial, using the notation
\begin{equation} \label{17}
{\mathcal P}\{\Delta(U_{T}-y) = 1\} = {\mathcal P}\{y
\leq U_{T}\} = p,
\end{equation}
we obtain
\begin{equation} \label{18}
{\mathcal P}\{ z_{N} = j\} = \binom{N}{j}\; p^{j}\;(1-p)^{N-j},
\;\;\;\;\;\;  j = 0, 1, \ldots, N.
\end{equation}

Our task is to find a confidence interval $[\gamma_{L}(k),\;\gamma
_{U}(k)]$ that covers the value $p$ with a prescribed probability
$\beta$ provided we have a sample of size $N$ and in that sample
$z_{N}=k \leq N$. The probability (\ref{17}) gives the probability
that the output $y$ is not larger than the technological limit
$U_{T}$. When the lower level $\gamma_{L}(k)$ of the confidence
interval is close to unity, we can claim at least with probability
$\beta$ that the chance of finding the output $y$ smaller than
$U_{T}$ is also close to unity and the system under consideration
can be regarded as safe at the level $[\beta \vert
\gamma_{L}(k)]$.

If the sample size $N > 50$, the random variable
\begin{equation} \label{19}
\frac{k - Np}{\sqrt{Np\;(1 - p)}} = \zeta_{k}
\end{equation}
has approximately normal distribution. Here $k$ is the number of
sample elements not exceeding $U_{T}$. Let $\beta$ denote the
confidence level, then
\[ {\mathcal P}\{\vert \zeta_{k} \vert \leq u_{\beta}\} =
2\Phi(u_{\beta}) - 1 = \beta, \] where $\Phi(x)$ is the standard normal distribution
function. This equation can be rewritten in the form
\[ {\mathcal P}\{\vert \zeta_{k} \vert \leq u_{\beta}\} =
{\mathcal P}\{ (N+u_{\beta}^{2})(p-\gamma_{L})(p-\gamma_{U}) \leq
0\} = \beta,\] where
\[ \gamma_{L} = \gamma_{L}(k, u_{\beta}) = \]
\begin{equation}\label{20}
= \frac{1}{N+u_{\beta}^{2}}\;\left[k + \frac{1}{2}u_{\beta}^{2} -
u_{\beta}\sqrt{k(1-k/N) + u_{\beta}^{2}/4}\right],
\end{equation}
and
\[ \gamma_{U} = \gamma_{U}(k, u_{\beta}) = \]
\begin{equation}\label{21}
= \frac{1}{N+u_{\beta}^{2}}\;\left[k + \frac{1}{2}u_{\beta}^{2} +
u_{\beta}\sqrt{k(1-k/N) + u_{\beta}^{2}/4}\right].
\end{equation}
Here $u_{\beta}$ is the root of
\[ \Phi(u_{\beta}) = \frac{1}{2}(1 + \beta). \]

In a number of cases it suffices to know the probability of the
event $\{\gamma _{L}(k, v_{\beta}) \leq p\}$. Since $\zeta_{k}$
with $k$ fixed is a decreasing function of $p$, the events
$\{\zeta_{k } \leq  v_{\beta}\}$ and $\{\gamma _{L}(k, v_{\beta})
\leq p\}$ are equivalent, hence \[ {\mathcal P}\{ \zeta_{k} \leq
v_{\beta}\} = {\mathcal P}\{ \gamma_{L}(k, v_{\beta}) \leq p\} =
\Phi(v_{\beta}) = \beta.\] Consequently, the operation of a system
can be regarded safe if the parameter $p$ for all output variables
is covered by $[\gamma_{L}(k, v_{\beta }), 1]$ with a prescribed
probability $\beta$, provided that $\gamma_{L}(k, v_{\beta})$ is
close to unity.

\begin{table}[ht!]
\caption{\label{tab3} {\footnotesize Number  of successes $k$ in a
sample of size $N$}}
\begin{center}
\begin{tabular}{|c|c|c|c|c|c|c|c|c|c|c|c|} \hline
$k$ & 99 & 108 & 118 & 128 & 137 & 147 & 157 & 166 & 176 & 185 &
195 \\ \hline $N$ & 100 & 110 & 120 & 130 & 140 & 150 & 160 & 170
& 180 & 190 & 200  \\ \hline
\end{tabular}
\end{center}
\end{table}

Table 3 gives the number of successes $k$ in a sample of size $N$
needed for acceptance at the level $\beta = \gamma_{L} = 0.95$. We
utilized approximate formula (\ref{20}) to derive the entries in
Tab. 3.

When the sample size is less than $50$, we may not apply the
asymptotically valid normal distribution. The below given
derivation of the confidence limits is a modified method proposed
by Clopper and Pearson \cite{clo34}. The probability of at least
$k$ successes from $N$ observations is given by
\[ S_{k}^{(N)}(p) = \sum_{j=0}^{k} \binom{N}{j}\; p^{j}\;(1-p)^{N-j},\] where $p =
{\mathcal P}\{y \leq U_{T}\}$. This formula can be recast as
\[ S_{k}^{(N)}(p) =
\frac{N!}{k!\;(N-k-1)!}\;\int_{0}^{1-p} (1-v)^{k}\;v^{N-k-1}\;dv,\] and it is clear from
that expression that $S_{k}^{(N)}(p)$ is a monotonously decreasing function of $p$.
Since
\[ S_{k}^{(N)}(p) = \left\{ \begin{array}{ll}
1, & \mbox{if $p=0$,} \\
\mbox{} & \mbox{} \\
0, & \mbox{if $p=1$,} \end{array} \right. \]  it assumes an
arbitrary value only once in the interval [0,1]. Consequently, a
$p = p_{\delta}$ value exists so that
\[ S_{k}^{(N)}(p_{\delta}) = \delta, \;\;\;\;\;\; \forall \; 0
< \delta < 1. \] Exploiting the monotony, we can construct a
function such that
\[  R_{k}^{(N)}(p) < R_{k}^{(N)}(p_{\delta}) = \delta,\]
when $p > p_{\delta}$. Such a function is
\[ R_{k}^{(N)}(p) = 1 - S_{k-1}^{(N)}(p) = \sum_{j=k}^{N}
\binom{N}{j}\; p^{j}\;(1-p)^{N-j}, \] Finally, we establish the
upper limit $\gamma_{U}$ from \[ S_{k}^{(N)}(\gamma_{U}) \leq
\frac{1}{2}(1-\beta),\] and the lower limit $\gamma _{L}$ from
\[ R_{k}^{(N)}(\gamma_{L}) \leq \frac{1}{2}(1-\beta). \]
\begin{figure} [ht!]
\protect \centering{
\includegraphics[height=5cm, width=7.5cm]{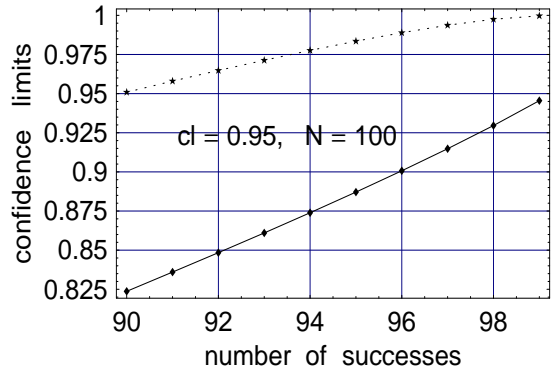}}
\caption{\label{fig2}{\footnotesize Dependence of $\gamma_{L}$ and
$\gamma_{U}$ on the number of successes in a sample of N=100
elements.}}
\end{figure}
The interval $[\gamma_{L}, \; \gamma_{U}]$  covers the unknown
parameter $p$ with probability $\beta $. The dependence of $\gamma
_{L}$ and $\gamma _{U}$ are shown in Fig. \ref{fig2} for a sample
of $N=100$ elements, $cl$ stands for confidence level $\beta$.

\subsection{Several output variables}

Now we assume the output to comprise $n$ variables. Let these variables be $y_{1},
\ldots, y_{n}$. There are several fairly good tests to prove if they are statistically
independent. To independent variables we can apply the considerations above but for
dependent variables we need novel considerations. Let
\[ \mathbf{S}_{N} = \left(\begin{array}{cccc}
y_{11} & y_{12} & \ldots & y_{1N} \\
y_{21} & y_{22} & \ldots & y_{2N} \\
\vdots & \vdots & \ddots & \vdots \\
y_{n1} & y_{n2} & \ldots & y_{nN}
\end{array} \right) \]
denote the sample matrix obtained in $N >> 2n$ independent
observations. With a computer model, an observation is a run.
Introducing the column vector $\vec{y}_{k}$, the sample matrix is
written as
\[ \mathbf{S}_{N} = \left(\vec{y}_{1}, \ldots, \vec{
y}_{N}\right). \]

Below we expound the sign test for two output variables $y_{1}$
and $y_{2}$ relying on the assumption that their joint
distribution function $G(y_{1}$,$y_{2})$ is unknown but continuous
in either variable. The goal of the foregoing analysis is to
verify the safety conditions  $y_{1} < U_{T}^{(1)}$ and $y_{2} <
U_{T}^{(2)}$. When the condition is accomplished with probability
$p_{12} = G(U_{T}^{(1)}, U_{T}^{(2)}) \approx 1$ we say the system
is safe. Here, as before, the limits $U_{T}^{(1)}$, and
$U_{T}^{(2)}$ are determined by the technology. Since $p_{12}$ is
unknown, our job is to construct a confidence interval
$[\gamma_{L}^{(1,2)},\; \gamma_{U}^{(1,2)}]$ so that it covers
$p_{12}$ with probability $\beta_{12}$. In most cases it suffices
to calculate solely $\gamma_{L}^{(1,2)}$ and to use the interval
$[\gamma_{L}^{(1,2)},\;1]$ as confidence interval. Now the column
vectors introduced above have two components. In accordance with
our assumption, different vectors are statistically independent
but the components in a given vector are not necessarily
independent. In order to keep the notation as simple as possible,
the event $\{y_{1} < U_{T}^{(1)},\; y_{2} < U_{T}^{(2)}\}$ will be
called a success. If $y_{1k} < U_{T}^{(1)}$ and $y_{2k} <
U_{T}^{(2)}$, then
\[ \Delta(U_{T}^{(1)}-y_{1k})\;\Delta(U_{T}^{(2)}-y_{2k})
= 1,\]  while $0$ otherwise, and introduce the statistical
function
\[ z_{N}^{(1,2)} = \sum_{k=1}^{N}
\Delta(U_{T}^{(1)}-y_{1k})\;\Delta(U_{T}^{(2)}-y_{2k})
\]
which gives the number of successes in the sample of size $N$.
Since the newly introduced random variable is the sum of $N$
independent random variables, assuming values either 1 or 0, its
distribution is binomial. Using the notation
\[ {\mathcal P}\{\Delta(U_{T}^{(1)}-y_{1})\;
\Delta(U_{T}^{(2)}-y_{2})=1\} = \]
\[ = {\mathcal P}\{y_{1} < U_{T}^{(1)}, \; y_{2} < U_{T}^{(2)}\}
= p_{12},  \] we can write
\[ {\mathcal P} \{z_{N}^{(1,2)}=k \} = \binom{N}{k}\;
p_{12}^{k}\;(1-p_{12})^{N-k},\] for $k=0,1, \ldots, N$. At this
point we rejoin the thought of line of the previous subsection.
Instead of repeating the already familiar argumentation, we amend
two trivial although important remarks. Let us define the
following two statistical functions:
\[ z_{N}^{(1)} = \sum_{i=1}^{N} \Delta(U_{T}^{(1)} -
y_{1i}) \] and  \[z_{N}^{(2)} = \sum_{j=1}^{N} \Delta(U_{T}^{(2)}
- y_{2j}). \] These two functions are not statistically
independent, either one is the sum of $N$ independent random
variables with values 1 or 0, therefore, one can write
\[ {\mathcal P}\{z_{N}^{(1)}=i\} = \binom{N}{i}\; p_{1}^{i}
(1-p_{1})^{N-i} \] and
\[ {\mathcal P}\{z_{N}^{(2)}=j\} = \binom{N}{j}\; p_{2}^{j}
(1-p_{2})^{N-j},  \]
\[ i,j = 1, \ldots, N, \]
where
\[ p_{\ell} = {\mathcal P}\{y_{\ell} < U_{T}^{(\ell)}\} =
{\mathcal P}\{\Delta(U_{T}^{(\ell)} - y_{\ell}) = 1\},
\] \[ \ell = 1, 2, \]
are unknown probabilities. Applying the method used previously,
this time separately to the samples  \[ {\mathcal S}_{N}^{(1)} =
\{y_{1i}, \;\; i=1, \ldots, N\}\] and \[{\mathcal S}_{N}^{(2)} =
\{y_{2j}, \;\; j=1, \ldots, N\}\] we construct two random
intervals $[\gamma_{L}^{(1)}, \;1]$ and $[\gamma_{L}^{(2)}, \;1]$
covering $p_{1}$ and $p_{2}$ with probabilities $\beta_{1}$ and
$\beta_{2}$, respectively.

Obviously, it could occur that the levels $(\beta_{1}\vert
\gamma_{L}^{(1)})$ and $(\beta_{2}\vert \gamma_{L}^{(2)})$
corroborate the claim that samples ${\mathcal S}_{N}^{(1)}$ and
${\mathcal S}_{N}^{(2)}$ separately comply with safety
requirements. This does not mean that we would arrive at the same
conclusion from analyzing the two sets jointly. The reason is that
$y_{1}$ and $y_{2}$, the two output random variables are not
statistically independent. Hence, we should ascertain weather the
interval $[\gamma_{L}^{(1,2)}, \;1]$ covers the probability
$p_{12}$ with the pre-assigned probability $\beta_{12}$. Since
$\gamma_{L}^{(1,2)} \leq \min\{\gamma_{L}^{(1)},
\gamma_{L}^{(2)}\}$, $\gamma_{L}^{(1)}$ and $\gamma_{L}^{(2)}$
would not contain information sufficient to declare the system
safe. Decision on the safety, when two output variables are
subjected to limitations should go as follows. Firstly, we test
the hypothesis concerning dependence of the output variables
$y_{1}$ and $y_{2}$. If they are dependent, we should estimate the
probability of the event $\{y_{1} < U_{T}^{(1)},\; y_{2} <
U_{T}^{(2)}\}$. Solely if they are statistically independent
should we estimate the probability of events $\{y_{1} <
U_{T}^{(1)}\},\; \{y_{2} < U_{T}^{(2)}\}$ independently.

Finally, we mention that the generalization of the sign test to
$n>2$ output variables is straightforward, we have to use the
statistical function
\begin{equation} \label{22}
z_{N}^{(1,\ldots,n)} = \sum_{k=1}^{N}\;\prod_{j=1}^{n}
\Delta(U_{T}^{(j)}-y_{jk})
\end{equation}
to evaluate safety based on observation of $N$ samples of the $n$
output variables. In this manner we obtain the sum of $N$
independent random variables in expression (\ref{22}), and then,
the further steps will be the same as at the beginning of the
subsection.
\begin{figure} [ht!]
\centering{
\includegraphics[height=5cm, width=7.5cm]{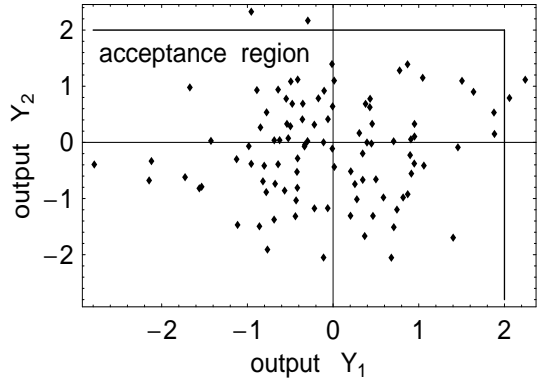}}
\caption{\label{fig3}{\footnotesize Sample a)}}
\end{figure}
\vspace{-0.1cm}
\begin{figure} [ht!]
\centering{
\includegraphics[height=5cm, width=7.5cm]{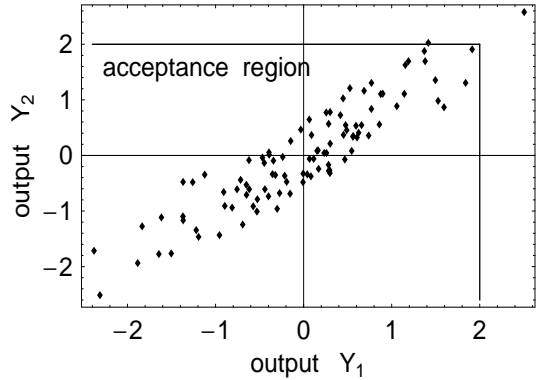}}
\caption{\label{fig4}{\footnotesize Sample b)}}
\end{figure}

An example is given below. We have generated two samples a) and b)
using Monte Carlo simulation, either sample contains $N=100$
observations (or runs) of two output variables. The samples have
been generated from a bivariate normal distribution with parameters
$m_{1}=m_{2}=0$ and $\sigma_{1}=\sigma_{2}=1$ but the correlation
coefficient is $C=0.1,\; C=0.7$ in sample a) and b), respectively.
The acceptance range is $[-2,2]$ for both output variables. In
sample a) and b) four and one samples lie respectively outside the
acceptance range. The results of the simulation can be seen in Fig.
\ref{fig3} and in Fig. \ref{fig4}.
\begin{table}[ht!]
\caption{\label{tab4}{\footnotesize Lower confidence limits in a
sample of $N=100$, $k$ is the number of success.}}
\begin{center}
\begin{tabular}{|c|c|c|c|} \hline
$k$ $\backslash$ $\beta$ & 0.90 & 0.95 & 0.99
\\ \hline 90 & 0.8501 & 0.8362 & 0.8086
\\ \hline 91 & 0.8616 & 0.8482 & 0.8212
\\ \hline 92 & 0.8733 & 0.8602 & 0.8340
\\ \hline 93 & 0.8850 & 0.8725 & 0.8471
\\ \hline 94 & 0.8970 & 0.8850 & 0.8604
\\ \hline 95 & 0.9092 & 0.8977 & 0.8741
\\ \hline 96 & 0.9216 & 0.9108 & 0.8882
\\ \hline 97 & 0.9344 & 0.9242 & 0.9030
\\ \hline 98 & 0.9476 & 0.9383 & 0.9185
\\ \hline 99 & 0.9616 & 0.9534 & 0.9354
\\ \hline 100 & 0.9772 & 0.9704 & 0.9549 \\ \hline
\end{tabular}
\end{center}
\end{table}

First let us consider sample a). From Tab. 4 one can read that the
interval [0.9108,1] covers the parameter $p_{12}$ with probability
$\beta_{12}$=0.95.

When we assess the output variables one by one, we see that the
associated probabilities $p_{1}$ and $p_{2}$ are covered by the
interval $[0.9383,1]$ with probability $\beta=0.95$ in either
sample. However tempting is to use $0.9383$ as lower bound for the
probability to be used in safety analysis, that number has nothing
to do with $p_{12}$ and should not be used in safety analysis.

Now let us pass on to sample b) where we see a strong correlation
between $y_{1}$ and $y_{2}$. From Tab. 4 one can read that the
confidence interval $[0.9383,1]$ covers the probability $\beta
_{12}=0.95$. From that sample we conclude that the probability of
the event $\{y_{1} < U_{T}^{(1)},\; y_{2} < U_{T}^{(2)}\}$ is at
least $0.9383$. The single variable parameters $p_{1}$ and $p_{2}$
determined from sample b) are covered by the intervals
$[0.9534,1]$ and $[0.9383,1]$, respectively on the level
$\beta_{1}=\beta_{2}=0.95$. Again, however favorable these numbers
are, they should not be used in assessing safety. The above
discussed simple numerical example clearly indicated the danger
awaiting the analyst when his/her judgment is based on tests
performed separately on correlated output variables.

\section{Concluding remarks}

The authors have investigated the statistical methods applied to
safety analysis of nuclear reactors and arrived at alarming
conclusions: Guba and Trosztel \cite{guba20} carried out a series
of calculations with the generally appreciated safety code ATHLET
to ascertain the stability of the results against input
uncertainties in a simple experimental situation. Scrutinizing
those calculations, we came to the conclusion [3] that the ATHLET
results may exhibit irregular  behavior. A further conclusion is
that the technological limits are incorrectly set [5] when the
output variables are correlated. Another formerly unnoticed
conclusion of the Guba-Trosztel calculations \cite{guba20} is that
certain innocent looking parameters (like wall roughness factor,
the number of bubbles per unit volume, the number of droplets per
unit volume) can influence considerably such output parameters as
water levels. The authors are concerned with the statistical
foundation of present day safety analysis practices and can only
hope that their own misjudgment will be dispelled.

Until then, the authors suggest applying correct statistical
methods in safety analysis even if it makes the analysis more
expensive. It would be desirable to continue exploring the role of
internal parameters (wall roughness factor, steam-water surface in
thermal hydraulics codes, homogenization methods in neutronics
codes) in system safety codes and to study their effects on the
analysis.

In the validation and verification process of a code one carries
out a series of computations. The input data are not precisely
determined because measured data have an error, calculated data
are often obtained from a more or less accurate model. Some users
of large codes are content with comparing the nominal output
obtained from the nominal input, whereas all the possible inputs
should be taken into account when judging safety. At the same
time, any statement concerning safety must be aleatory, and its
merit can be judged only when the probability is known with which
the statement is true. In some cases statistical aspects of safety
are misused as in \cite{krzy90}, where the number of runs for
several outputs is correct only for statistically independent
outputs, or misinterpreted as in \cite{wall03}.

We do not know the probability distribution of the output
variables subjected to safety limitations. At the same time in
some asymmetric distributions the $0.95 \vert 0.95$ methodology
simply fails: if we repeat the calculations in many cases we would
get a value higher than the basic value, which means the limit
violation in the calculation becomes more and more probable in the
repeated analysis.

Consequent application of order statistics or the application of
the sign test may offer a way out of the present situation. The
authors are also convinced that efforts should be made
\begin{itemize}
\item to study the statistics of the output variables,
\item to study the occurrence of chaos in the analyzed cases.
\end{itemize}

All these observations should influence, in safety analysis, the
application of best estimate methods, and underline the opinion
that any realistic modelling and simulation of complex systems
must include the probabilistic features of the system and the
environment.

\appendix
\section*{Appendix}

Let $\eta$ be a random variable with continuous distribution defined
over the real numbers ${\mathcal R}$, and  let the distribution
function of $\eta$ be
\begin{equation}\label{A1}
    \mathcal{P}\left\{\eta\le y \right\} = G(y).
\end{equation}
We carry out $N$ statistically independent observations of $\eta$.
That operation is called ${\mathcal K}$. We repeat ${\mathcal K}$
$n+1$ times. We group the observed values into the following
$(n+1)\times N$ matrix:
\begin{equation}\label{A2}
    \begin{array}{cccc}
      \eta_{01} & \eta_{02} & \cdots & \eta_{0N} \\
      \eta_{11} & \eta_{12} & \cdots & \eta_{1N} \\
      \dots & \dots & \dots & \cdots \\
      \eta_{n1} & \eta_{n2} & \cdots & \eta_{nN}
    \end{array} .
\end{equation}
Let denote $\zeta_j=\max_{1\le k\le N}{\eta_{jk}}$ the maximum
observed in operation $j$.
\par
{\bf Lemma.} Since the probability density function $G(y)$ is
monotonously increasing, and continuous, the following equation
holds for $0\le \gamma\le 1$:
\begin{equation}\label{A3}
    \mathcal{P}\left\{\max_{1\le k \le N}\eta_{jk}>G^{-1}(\gamma) \right\}=
    \mathcal{P}\left\{\int_{-\infty} ^{\zeta_j}dG(y)>\gamma
    \right\}=1-\gamma^N ,
\end{equation}
where $G^{-1}(\gamma)=Q_\gamma$ is the $\gamma$ quantile of the
probability density distribution function $G(y)$.
\par
The presented Lemma is well known, we omit its proof. Now we turn to the determination
of the probability distribution of the largest sample elements.
\par {\bf Theorem.}The probability of the event that among
the independent random variables $\zeta_1,\dots,\zeta_N$ there is
$k\le N$ greater than $\zeta_0$ is
\begin{equation}\label{A4}
    P_k=1-\frac{k}{n+1}.
\end{equation}
{\em Proof}: Since $\eta_{jk}, j=0, 1, \dots, n; k=1,\dots,N$ are
independent and identically distributed, we have
\begin{equation}\label{A5}
    \mathcal{P}\left\{\zeta_j\le z\right\}=\mathcal{P}\left\{\max_{1\le k\le
    N} \eta_{jk}\le z     \right\}=
    \prod_{k=1} ^N \mathcal{P}\left\{ \eta_{jk}\le z\right\}=H(z).
\end{equation}
In other words, $H(z)$ is the probability of $\zeta_j$ not being
larger than $z\in {\mathcal R}$ for any $j=0,1,\dots,n$. Let $0\le
\nu_n(z)\le n$ denote the number of those variables from among
$\zeta_1,\ldots,\zeta_n$ which are greater than $z$. Obviously,
\begin{equation}\label{A6}
\mathcal{P}\left\{ \nu_n(z)=\ell \right\}=J_\ell ^{(n)}(z) =
\binom{n}{\ell}\;\left( 1-H(z)\right) ^\ell \left(
H(z)\right)^{(n-\ell)} .
\end{equation}
Let $P_k$ stand for the probability that from among the random
variables $\zeta_1,\dots,\zeta_n$ at least $k\le n$ is greater than
$\zeta_0$, which may take any number from ${\mathcal R}$. We get
\begin{equation}\label{A7}
    P_k=\sum_{\ell=k} ^n p_\ell=\sum_{\ell=k} ^n \int_{-\infty} ^{+\infty}J_\ell
    ^{(n)}(z)dH(z).
\end{equation}
The determination of probabilities $p_\ell$ is straightforward:
\begin{equation}\label{A8}
    p_\ell=\int_{-\infty} ^{+\infty} J_\ell ^{(n)}(z)dH(z)=
    \binom{n}{\ell}\; \int_{-\infty} ^{+\infty} \left( 1-H(z)\right)^\ell \left(
    H(z)\right)^{n-\ell}dH(z).
\end{equation}
The integrals are evaluated without difficulties:
\begin{equation}\label{A9}
    p_\ell = \binom{n}{\ell} \;\int_0 ^1 (1-u)^\ell
    u^{n-\ell}du=\frac{1}{n+1}.
\end{equation}
As we see, $p_\ell$ is independent of $\ell$ and using Eq.
(\ref{A7}), we get
\begin{equation}\label{A10}
    P_k=\sum_{\ell=k} ^n
    \frac{1}{n+1}=\frac{n-k+1}{n+1}=1-\frac{k}{n+1}.
\end{equation}
Q.E.D.
\par
We add two remarks.
\begin{enumerate}
  \item {\bf Remark 1.} Whichever we choose from among the random variables
  $\zeta_0,\zeta_1,\dots,\zeta_n$, with probability $\frac{1}{n+1}$ we find among the others  $\ell$
  exceeding the first chosen one. (Since
  $\zeta_0,\zeta_1,\dots,\zeta_n$ are continuous random variables,
  the probability of two values to be identical is zero.)
  \item {\bf Remark 2.} Let $\lambda$ be the number of those
  $\zeta_{j1}, \zeta_{j2}, \ldots, \zeta_{jn}$ variables which are greater than a
  given $\zeta_{j0}$. Clearly, $\lambda$ is a random variable, its
  expectation value is
  \begin{equation}\label{A11}
    \mathbf{E}\left\{ \lambda \right\}=\sum_{\ell=0} ^n \ell
    p_\ell=\frac{n}{2},
  \end{equation}
the variance being
\begin{equation}\label{A12}
    \mathbf{D}^2\left\{ \lambda \right\}=\sum_{\ell=0} ^n
    \left(\ell-n/2\right)^2 p_\ell=\frac{1}{6}n\left(1+\frac{1}{2}n
    \right).
\end{equation}
\end{enumerate}

\end{document}